\documentclass[12pt,preprint]{aastex6}
\usepackage{graphicx,natbib,amsmath}
\usepackage{epstopdf}

\newcommand{\lapprox} {\, \lower3pt\hbox{$\sim$}\llap{\raise2pt\hbox{$<$}}\,}
\newcommand{\gapprox} {\, \lower3pt\hbox{$\sim$}\llap{\raise2pt\hbox{$>$}}\,}

\begin{document}

\title{ENERGY DEPOSITION BY ENERGETIC ELECTRONS IN A DIFFUSIVE COLLISIONAL TRANSPORT MODEL}

\author{A. Gordon Emslie\altaffilmark{1}, Nicolas H. Bian\altaffilmark{1,2}, and Eduard P. Kontar\altaffilmark{2}}

\altaffiltext{1}{Department of Physics \& Astronomy, Western Kentucky University, Bowling Green, KY 42101 (emslieg@wku.edu)}

\altaffiltext{2}{School of Physics \& Astronomy, University of Glasgow, Glasgow G12 8QQ, Scotland, UK}

\begin{abstract}

A considerable fraction of the energy in a solar flare is released as suprathermal electrons; such electrons play a major role in energy deposition in the ambient atmosphere and hence the atmospheric response to flare heating. Historically the transport of these particles has been approximated through a deterministic approach in which first-order secular energy loss to electrons in the ambient target is treated as the dominant effect, with second-order diffusive terms (in both energy and angle) being generally either treated as a small correction or neglected. However, it has recently been pointed out that while neglect of diffusion in energy may indeed be negligible, diffusion in angle is of the same order as deterministic scattering and hence must be included. Here we therefore investigate the effect of angular scattering on the energy deposition profile in the flaring atmosphere. A relatively simple compact expression for the spatial distribution of energy deposition into the ambient plasma is presented and compared with the corresponding deterministic result. For unidirectional injection there is a significant shift in heating from the lower corona to the upper corona; this shift is much smaller for isotropic injection.  We also compare the heating profiles due to return current Ohmic heating in the diffusional and deterministic models.

\end{abstract}

\keywords{acceleration of particles -- Sun: activity -- Sun: flares -- Sun: X-rays, gamma rays}

\section{Introduction}

Energy transport in solar flares involves a variety of mechanisms, such as nonthermal particle acceleration and propagation, thermal conduction, radiation, and bulk mass motions \citep[see, e.g.,][for reviews]{1988psf..book.....T,2011SSRv..159..107H,2011SSRv..159..301K}.
A significant fraction \citep[e.g.,][]{2012ApJ...759...71E}
of the energy released is manifested as bremsstrahlung-emitting deka-keV electrons \citep[see, e.g.,][]{1976SoPh...50..153L,2011SSRv..159..357Z}.
These electrons propagate from the primary energy release site and deposit their energy in the ambient target principally through Coulomb collisions on ambient electrons \citep[e.g.,][]{1972SoPh...26..441B,1978ApJ...224..241E}, with additional energy losses associated with Ohmic dissipation of the neutralizing return current \citep[e.g.,][]{1980ApJ...235.1055E,2006ApJ...651..553Z} and with the turbulent environment through which they propagate \citep[e.g.,][]{2011ApJ...730L..22K,2011A&A...535A..18B}.

Modeling of the Coulomb collision process has typically
involved a test-particle approach involving systematic (secular) energy loss \citep[e.g.,][]{1971SoPh...18..489B,1972SoPh...26..441B,1978ApJ...224..241E},
although numerical solutions of the Fokker-Planck equation, involving collisional diffusion in pitch angle \citep[e.g.][]{1981ApJ...251..781L,1991A&A...251..693M,1991ApJ...374..369B,2014ApJ...780..176K} and energy \citep[e.g.][]{2014ApJ...787...86J}
in addition to the secular energy loss term have also been carried out. \cite{2017ApJ...835..262B} have shown that, while diffusion in {\it energy} can be justifiably neglected in a sufficiently cold target, diffusion of the accelerated electrons in {\it angle} is of the same order as secular change in angle and thus it is essential to include diffusive angular scattering processes in determining the spatial and angular distributions of the accelerated electrons in the target.

Knowledge of the energy deposition profile is a key element in determining the response of the solar atmosphere to flare heating \citep{1989ApJ...341.1067M,2015ApJ...809..104A} and
hence in interpreting the plethora of observations of Doppler-shifted and -broadened spectral lines \citep[e.g.,][]{1982SoPh...78..107A,1987SoPh..110..295E,1990ApJS...73..117D,1993ApJ...419..418M,1994ApJ...421L..55B,2001ApJ...554..464R,2003ApJ...582..506L,2004ApJ...613..580B,2015ApJ...813...59L,2015ApJ...811....7L,2015SoPh..290.3399M,2015ApJ...808..177R,2015ApJ...811..139T,2016ApJ...830..101B,2017ApJ...848...39B,2016A&A...588A...6G,2016ApJ...816...89P,2016ApJ...829...35W,2017PhRvL.118o5101K,2017ApJ...841L...9L} in terms of the velocity differential emission measure \citep{1995ApJ...447..915N} corresponding to candidate energy transport models. In this paper we therefore build on the results of \citet{2017ApJ...835..262B} to derive a formula for the energy deposition profile associated with the passage of electrons through a cold target, where diffusion associated with angular scattering is explicitly taken into account.  The results show that for unidirectional injection the spatial distribution
of plasma heating differs noticeably from the simple deterministic treatment that has formed the basis for much of the modeling of both solar \citep{1973SoPh...31..143B,1989ApJ...341.1067M} and
stellar \citep{2015ApJ...809..104A} flares to date.

In Section~\ref{diffusive-solutions} we present an analysis of collision-dominated electron propagation in a cold target, with angular diffusion taken into account;
the results are presented as a solution for the electron flux $F(E,z)$ (electrons~cm$^{-2}$~s$^{-1}$~keV$^{-1}$) at energy $E$ and target depth $z$ in terms of an integral over a Green's function for electrons injected at a specified energy and pitch angle. In Section~\ref{energy-deposition-rates} we use this result to calculate the energy deposition rate as a function of $z$, both for unidirectional and isotropic injection cases. In Section~\ref{return-current-ohmic} we briefly discuss the impact of diffusive angular scattering on the return current Ohmic losses associated with driving the beam-neutralizing electron current though the finite resistivity of the ambient plasma. In Section~\ref{summary-conclusions} we discuss the results and present our conclusions.

\section{Solution to the collisional transport equation in the diffusive regime}\label{diffusive-solutions}

\cite{2017ApJ...835..262B} have shown that the collisional transport of electrons in a cold target can effectively be modeled, in a first (local)
approximation\footnote{In general, as shown by \citet{2017ApJ...835..262B}, the diffusive term is non-local, so that the corresponding particle flux depends on the spatial gradient of the electron distribution function over a range of distances $\sim \lambda/\sqrt{45}$, where $\lambda$ is the collisional mean free path. We neglect this higher-order effect in this work.}, by the one-dimensional transport equation \citep[e.g.][]{2014ApJ...780..176K}

\begin{equation}\label{cold-diffusion}
- \, \frac{\partial }{\partial z} \left ( \frac{\lambda_{C}(v)v}{6} \, \frac{\partial  f_{0}(z,v)}{\partial z} \right ) =  \frac{1}{v^{2}} \, \frac{\partial}{\partial v} \left ( v^{3} \, \nu_{C}(v) \, f_{0} \right ) + S_{0}(z,v) \,\,\, .
\end{equation}
Here $f_0(v,z)$ (electrons~cm$^{-3}$~[cm~s$^{-1}$]$^{-3}$)
is the principal (isotropic) part of the electron phase space distribution at speed $v$ and distance $z$ from the injection site, $S_0(z,v)$ (electrons~cm$^{-3}$~s$^{-1}$~[cm~s$^{-1}$]$^{-3}$)
is the injection (source) term, and the collisional mean-free path

\begin{equation}\label{lambda_C}
\lambda_{C}(v) = \frac{v}{\nu_{C}(v)} \,\,\, ,
\end{equation}
with $\nu_C(v)$ the cold-target collision frequency, given by

\begin{equation}\label{nu-c}
\nu_{C}(v) = \frac{4\pi n_e \, e^4 \, \ln \Lambda}{m_e^2} \, \frac{1}{v^3} \,\,\, .
\end{equation}
In this equation $n_e = 4 \pi \int f_0(v) \, v^2 \, dv$ is the local density (cm$^{-3}$), $e$ (esu) and $m_e$ (g) are the electronic charge and mass, respectively, and $\ln \Lambda$ is the Coulomb logarithm \citep[e.g.,][]{1962pfig.book.....S}.

It is convenient to make a transformation of the dependent variable
from $f_0(v,z)$ to the energy flux $F(E,z)$
(electrons~cm$^{-2}$~s$^{-1}$~erg$^{-1}$).
This is related to the phase-space distribution function $f_0$ (electrons~cm$^{-3}$~[cm~s$^{-1}]^{-3}$) by considering
the hemispherical particle flux, i.e.,

\begin{equation}
F(E,z) \, dE = f_{0}(v,z) \, v^2 \, dv \int_{\phi = 0}^{2 \pi} \int_{\theta =0}^{\pi/2} v \cos \theta \, \sin \theta \, d\theta \, d\phi = \pi f_0 \, v^3 \, dv  \,\,\, .
\end{equation}
Using $dE= mv \, dv$ we obtain the relation

\begin{equation}\label{F-f}
F(E,z) = \frac{\pi}{m_e} \, f_0 (v,z) \, v^2 \,\,\, .
\end{equation}
Substituting Equations~(\ref{lambda_C}), (\ref{nu-c}),
and~(\ref{F-f}) in Equation~(\ref{cold-diffusion}),
we obtain the diffusion equation

\begin{equation}\label{diffusion-continuity}
- \, \frac{\lambda_C(E)}{6} \, \frac{\partial^2 F(E,z)}{\partial z^2} + \frac{\partial }{\partial E} \, \left [ \, B(E) \, F(E,z) \, \right ] = {\hat S}(E,z) \,\,\, ,
\end{equation}
where ${\hat S}(E,z)$ (cm$^{-3}$~s$^{-1}$~erg$^{-1}$) = $(\pi v/m_e) \, S_0(v,z)$, $\lambda_C(E)$ is the collisional mean free path as a function of energy $E$ (cf. Equations~(\ref{lambda_C}) and~(\ref{nu-c})):

\begin{equation}\label{lambdae-def}
\lambda_C(E)= \frac{E^2}{\pi n_e e^4 \ln \Lambda} \equiv \frac{2 E^2}{Kn} \,\,\, ,
\end{equation}
and $B(E)$ (erg~cm$^{-1}$) is the usual \citep{1972SoPh...26..441B,1978ApJ...224..241E} cold-target energy loss rate per unit distance:

\begin{equation}\label{cold-target-be}
B(E) \equiv \frac{dE}{dz} = - \frac{2\pi e^4 \ln \Lambda \, n}{E} \equiv -\frac{Kn}{E} \,\,\, .
\end{equation}
Equation~(\ref{cold-target-be}) has solution $E^2(z) = E^2(0) - 2K \int n(z) \, dz$.  For simplicity, we shall henceforth assume a uniform density $n$, so that $E^2= E^2(0) - 2Knz$.  The characteristic collisional stopping distance, for an electron of injected energy $E$ in a scenario without diffusion, is $E^2/2Kn$, is thus one-fourth of the diffusional mean free path~(\ref{lambdae-def}).

We now change to the new dependent variable

\begin{equation}\label{Phi-def}
\Phi(E,z) = F(E,z) \, B(E) \equiv -\frac{Kn}{E} \, F(E,z)
\end{equation}
(units cm$^{-3}$~s$^{-1}$) and to a new independent energy variable (with units cm$^2$)

\begin{eqnarray}\label{zeta-def}
\zeta = \frac{1}{6} \, \int dE \, \frac{\lambda_C(E)}{B(E)} & = & - \frac{1}{6Kn} \, \int_{E_{0}}^{E}  \lambda_C(E') \, E' \, dE' = \frac{1}{3 (Kn)^2} \int_E^{E_{0}} E'^3 \, dE' \cr
&=& \ell^{2} \left [ \left ( \frac{E_{0}}{k_{B}T_{e}} \right )^{4} -
\left (\frac{E}{k_{B}T_{e}} \right )^{4} \right ] \,\,\, ,
\end{eqnarray}
where

\begin{equation}\label{ell-def}
\ell = \frac{1}{2 \sqrt{3}} \frac{(k_B T_e)^2}{Kn} \equiv \frac{\lambda_{ec}}{4 \sqrt{3}} ; \qquad \lambda_{ec} = \frac{2 (k_B T_e)^2}{K \, n} \,\,\, .
\end{equation}
With this substitution, Equation~(\ref{diffusion-continuity}) takes the form of a standard diffusion equation

\begin{equation}\label{basic}
\frac{\partial \Phi}{\partial \zeta} = \frac{\partial^2 \Phi(E,z)}{\partial z^2} + {\overline S}(\zeta,z) \,\,\, ,
\end{equation}
where ${\overline S}(\zeta,z) = (6 B(E)/\lambda_C(E)) \, {\hat S}$ (cm$^{-5}$~s$^{-1}$) is the pertinent source function. The well-known Green's function for such a parabolic diffusion equation is

\begin{equation}
G_{\Phi}(\zeta,z) = \frac{1}{(4\pi \zeta)^{1/2}} \, \exp \left ( - \frac{({z-z'})^{2}}{4\zeta} \right ) \,\,\, ,
\end{equation}
or, in terms of the original independent variables $(E,z)$ and dependent variable $F(E,z)$,

\begin{equation}
G_{F}(E,z) = \frac{E}{Kn \left \{ 4\pi l^{2} \left [ \left (\frac{E_{0}}{k_{B}T_{e}} \right )^{4}-
\left ( \frac{E}{k_{B}T_{e}} \right )^{4} \right ] \right \}^{1/2}} \, \exp \left \{ - \, \frac{({z-z'})^{2}}{4 l^{2} \left [ \left ( \frac{E_{0}}{k_{B}T_{e}} \right )^{4}-
\left ( \frac{E}{k_{B}T_{e}} \right )^{4} \right ]} \right \} \,\,\, .
\end{equation}
Hence, the solution to Equation~(\ref{diffusion-continuity}), with a source term of the form ${\hat S}(E,z) = S(z) \, F_{0}(E_0)$, where $S(z)$ has units cm$^{-1}$ and $F_0(E_0)$ has units cm$^{-2}$~s$^{-1}$~erg$^{-1}$, can be expressed
\citep[see Eq. (26) in][]{2014ApJ...780..176K} as

\begin{eqnarray}\label{f-result}
F(E,z) & = & \frac{E}{Kn}\int _{-\infty}^{+\infty}dz'\int_{E}^{\infty} dE_{0}
\frac{S(z') \, F_{0}(E_{0})}{\left \{ 4\pi l^{2} \left [ \left (\frac{E_{0}}{k_{B}T_{e}} \right )^{4}-
\left ( \frac{E}{k_{B}T_{e}} \right )^{4} \right ] \right \}^{1/2}} \, \times \cr
& \times & \, \exp \left \{ - \, \frac{({z-z'})^{2}}{4 l^{2} \left [ \left ( \frac{E_{0}}{k_{B}T_{E}} \right )^{4}-
\left ( \frac{E}{k_{B}T_{E}} \right )^{4} \right ]} \right \} \,\,\, .
\end{eqnarray}

To illustrate the form of this solution, and in particular how it deviates from the diffusion-free result of past works, let us assume for definiteness a point-injection

\begin{equation}
S(z)=\delta(z)
\end{equation}
and a low-energy-truncated power-law injection form for the source (acceleration) spectrum:

\begin{equation}
F_{0}(E_0)=\frac{\dot{N}}{A} \, \frac{(\delta -1)}{E_{c}} \, \left ( \frac{E_0}{E_{c}} \right )^{-\delta} \, H(E_0 - E_c) \,\,\, ,
\end{equation}
where $H(x)$ is the Heaviside step function and the total injected rate (s$^{-1}$)

\begin{equation}
\dot{N} = A \, \int_{E_{c}}^{\infty} F_0(E_0) \, dE_0 \,\,\, .
\end{equation}
With these identifications, we obtain

\begin{equation}\label{fez-result}
F(E,z) = \sqrt{\frac{3}{\pi}} \, \frac{\dot{N}}{A} \,
\frac{(\delta-1)}{E_{c}} \times
\begin{cases}
E \, \int_{E}^{\infty} dE_{0} \, \frac{(E_{0}/E_{c})^{-\delta}}{(E_0^4-E^4)^{1/2}} \, \exp \left \{ - \frac{3 (Knz)^2}{E_0^4-E^4} \right \} \qquad E \ge E_c \cr
E \, \int_{E_c}^{\infty} dE_{0} \, \frac{(E_{0}/E_{c})^{-\delta}}{(E_0^4-E^4)^{1/2}} \, \exp \left \{ - \frac{3 (Knz)^2}{E_0^4-E^4} \right \} \qquad E < E_c \,\,\, .
\end{cases}
\end{equation}

We can compare this expression with that for one-dimensional deterministic transport.  Unlike for the diffusional case\footnote{in the original Fokker-Planck equation we can write only $d<\!\!\! E \!\!\! >/dz=-Kn/z$, which does {\it not} allow a deterministic relation between $E$ and $z$.}, there now {\it is} a unique value of the energy $E$ at position $z$.  For a one-dimensional transport model, this is given by

\begin{equation}\label{ee0}
E^2 = E_0^2 - 2 K n z \,\,\, .
\end{equation}
Further, since all the energy is injected in one direction, we need consider only $z \ge 0$.  The corresponding expression for $F(E,z)$ is \citep[e.g.,][]{1984ApJ...279..882E}

\begin{eqnarray}\label{fez-old-model}
F_{ND}(E,z) & = & F_0(E_0) \, \frac{dE_0}{dE} = \frac{E}{E_0} \, F_0(E_0) = \cr
& = & \begin{cases}
\frac{\dot{N}}{A} \,
(\delta-1) \, E_c^{\delta-1} \, \frac{E}{\left ( E^2 + 2Knz \right )^{(\delta+1)/2} } \quad ; \quad E^2 \ge E_c^2 - 2 K n z \\
0 \qquad \qquad \qquad \qquad \qquad \qquad \quad \,\,
; \quad {\rm otherwise} \,\,\, .
\end{cases}
\end{eqnarray}

\section{Collisional energy deposition rate}\label{energy-deposition-rates}

With the forms of $F(E,z)$ now determined, we turn our attention to the energy deposition profile due to Coulomb collisions, thus generalizing the diffusionless treatments of \cite{1973SoPh...31..143B} and \cite{1978ApJ...224..241E}. We remind the reader that even in the diffusional model, the diffusion is in pitch angle only \citep[diffusion in energy is a higher order effect;][]{2017ApJ...835..262B}, so that a cold-target energy loss rate $dE/dz = -Kn/E$ is still appropriate for each electron.

\subsection{Non-diffusional model}\label{edep-nd}

We first review the results for the deterministic non-diffusional model.  Although these results are well established in the literature, dating back to \cite{1972SoPh...26..441B,1973SoPh...31..143B}, it is worth reviewing these to provide a baseline and also to develop a method that carries over to the diffusive case.

In the non-diffusive case, the heating rate $Q(z)$ can be obtained by evaluating \citep[cf.][]{1973SoPh...31..143B,1978ApJ...224..241E} the quantity

\begin{equation}\label{qnd-alternative}
Q(z) = \int_0^\infty F_{ND}(E,z) \, \left \vert \frac{dE}{dz} \right \vert \, dE = Kn \int_0^\infty \frac{F_{ND}(E,z)}{E} \, dE \,\,\, .
\end{equation}
Substituting for $F_{ND}(E,z)$ from Equation~(\ref{fez-old-model}), we obtain

\begin{eqnarray}\label{q-nd-de-dz-method}
Q(z) = \frac{{\dot N}}{A} \, (\delta-1) E_c^{\delta-1} Kn \times \begin{cases}
\int_{\sqrt{E_c^2-2Knz}}^\infty \frac{dE}{\left ( E^2 + 2Knz \right )^{(\delta+1)/2}} \quad ; \quad z < \frac{E_c^2}{2Kn} \\
\int_0^\infty \frac{dE}{( E^2 + 2Knz)^{(\delta+1)/2}} \qquad \qquad \, ; \quad z > \frac{E_c^2}{2Kn} \,\,\, .
\end{cases}
\end{eqnarray}
Using the substitution

\begin{equation}\label{y-variable}
y = \frac{2Knz}{E^2+2Knz} \,\,\, ,
\end{equation}
Equation~(\ref{q-nd-de-dz-method}) can be written as

\begin{equation}\label{q-nd-de-dz}
Q(z) = \frac{1}{2} \, (\delta-1) \, \frac{{\dot N}}{A} \, \frac{Kn}{E_c} \times
\begin{cases}
B_{(2Knz/E_c^2)} \left ( \frac{\delta}{2}, \frac{1}{2} \right ) \, \left ( \frac{2Knz}{E_c^2} \right )^{-\delta/2} \,\, ; \quad z < \frac{E_c^2}{2Kn} \\
B \left ( \frac{\delta}{2}, \frac{1}{2} \right ) \, \left ( \frac{2Knz}{E_c^2} \right )^{-\delta/2} \qquad \qquad ; \quad z > \frac{E_c^2}{2Kn} \,\,\, ,
\end{cases}
\end{equation}
where the incomplete beta function is

\begin{equation}\label{incomplete-beta}
B_x(a,b) = \int_0^x y^{a-1} \, (1-y)^{b-1} \, dy
\end{equation}
and the complete beta function $B(a,b) \equiv B_1 (a,b)$.

For injection at an angle to the guiding magnetic field, the electrons propagate through the target with varying pitch angle, and the relationship between the energy and pitch angle at a given depth to the injected energy and pitch angle is more complicated \citep{1972SoPh...26..441B}.  The corresponding heating rate can, however, be well approximated as a straightforward generalization of Equation~(\ref{qnd-alternative}), namely

\begin{equation}\label{qnd-general-anisotropic}
Q_{ND}(z) = Kn \int_{\mu=0}^1 h(\mu) \int_{0}^{\infty} \frac{F(E,z/\mu)}{\mu \, E} \,\, dE \, d\mu \,\,\, ,
\end{equation}
where $h(\mu) \, d\mu$ is the fraction of the flux at pitch angle cosines in $(\mu, \mu + d\mu)$.  In particular, if the electrons are injected isotropically in the half-plane, then \citep[see Equation~(25) of ][]{1972SoPh...26..441B} they remain isotropic at all depths, and

\begin{equation}\label{qnd-general-isotropic}
Q_{ND}(z) = Kn \int_{\mu=0}^1 \int_{0}^{\infty} \frac{F(E,z/\mu)}{\mu \, E} \,\, dE \, d\mu \,\,\, .
\end{equation}
This expression can be readily evaluated numerically using the obvious generalization of Equation~(\ref{q-nd-de-dz}).

Because in the diffusive case there is no unique value of $E$ associated with an electron injected with energy $E_0$ at position $z$, the above expression (or a generalization of it) cannot be used.  We therefore develop an expression for the heating rate that can also be applied to the diffusive case. We first use Equation~(\ref{fez-old-model}) to obtain an expression for the total energy flux ${\cal F}(z)$ (erg~cm$^{-2}$~s$^{-1}$) at point $z$:

\begin{equation}\label{calf-z}
{\cal F}(z) = \int_0^\infty E \, F(E,z) \, dE =
\frac{{\dot N}}{A} \, (\delta-1) E_c^{\delta-1} \times \begin{cases}
\int_{\sqrt{E_c^2-2Knz}}^\infty \frac{E^2 \, dE}{\left ( E^2 + 2Knz \right )^{(\delta+1)/2}} \, ; \quad z < \frac{E_c^2}{2Kn} \\
\int_0^\infty \frac{E^2 \, dE}{( E^2 + 2Knz)^{(\delta+1)/2}} \qquad \quad \,\,
; \quad z > \frac{E_c^2}{2Kn} \,\,\, .
\end{cases}
\end{equation}
Using the change of variable~(\ref{y-variable}), this can be written as

\begin{equation}\label{calf-z-alt}
{\cal F}(z) = \frac{1}{2} \, (\delta -1) \, \frac{{\dot N}}{A} \, E_c \, \left (  \frac{2 K n z}{E_c^2} \right )^{1-\delta/2} \times
\begin{cases}
B_{(2Knz/E_c^2)} \left ( \frac{\delta}{2}-1, \frac{3}{2} \right )  \, \qquad \qquad \qquad ; \quad z < \frac{E_c^2}{2Kn} \\
B \left ( \frac{\delta}{2} - 1, \frac{3}{2} \right )  \, \qquad \qquad \qquad \qquad \qquad ; \quad z > \frac{E_c^2}{2Kn} \,\,\, .
\end{cases}
\end{equation}

Energy conservation requires that the heating rate $Q(z)$ (erg~cm$^{-3}$~s$^{-1}$) is

\begin{eqnarray}\label{q-nd-df-dz}
Q(z) & = & - \frac{d{\cal F}}{dz} =
\frac{{\dot N}}{A} \, (\delta-1) \, (\delta+1) \, Kn \, E_c^{\delta -1 } \times \cr
& \times &
\begin{cases}
\int_{\sqrt{E_c^2-2Knz}}^\infty \frac{E^2 \, dE}{ \left ( E^2+2Knz \right )^{(\delta+3)/2}} - \frac{\sqrt{E_c^2 - 2Knz}}{E_c} \, ; \quad z < \frac{E_c^2}{2Kn} \\
\int_0^\infty \frac{E^2 \, dE}{ \left ( E^2+2Knz \right )^{(\delta+3)/2}} \, \qquad \qquad \qquad \qquad \quad ; \quad z > \frac{E_c^2}{2Kn} \,\,\, ,
\end{cases}
\end{eqnarray}
which, using the change of variable~(\ref{y-variable}), can be written as

\begin{eqnarray}\label{q-nd}
Q(z) & = & \frac{1}{2} \, (\delta-1) \, (\delta+1) \,  \frac{{\dot N}}{A} \, \frac{Kn}{E_c} \times \cr
& \times &
\begin{cases}
B_{(2Knz/E_c^2)} \left ( \frac{\delta}{2}, \frac{3}{2} \right ) \, \left ( \frac{2Knz}{E_c^2} \right )^{-\delta/2}  - 2 \, \sqrt{1 - \frac{2Knz}{E_c^2}}\, ; \quad z < \frac{E_c^2}{2Kn} \\
B \left ( \frac{\delta}{2}, \frac{3}{2} \right ) \, \left ( \frac{2Knz}{E_c^2} \right )^{-\delta/2} \, \qquad \qquad \qquad \qquad \qquad \quad ; \quad z > \frac{E_c^2}{2Kn} \,\,\, .
\end{cases}
\end{eqnarray}
To simplify this expression, we note that

\begin{equation}\label{beta-identity}
(\delta+1) B \left ( \frac{\delta}{2}, \frac{3}{2} \right ) \equiv (\delta + 1) \frac{\Gamma \left ( \frac{\delta}{2} \right ) \Gamma \left ( \frac{3}{2} \right )}{\Gamma \left ( \frac{\delta+3}{2} \right )} = (\delta + 1) \frac{\Gamma \left ( \frac{\delta}{2} \right ) \frac{1}{2} \Gamma \left ( \frac{1}{2} \right )}{\left ( \frac{\delta+1}{2} \right ) \Gamma \left ( \frac{\delta+1}{2} \right )} = \frac{\Gamma \left ( \frac{\delta}{2} \right ) \Gamma \left ( \frac{1}{2} \right )}{\Gamma \left ( \frac{\delta+1}{2} \right )} \equiv B \left ( \frac{\delta}{2}, \frac{1}{2} \right ) \,\,\, ,
\end{equation}
and using this in the incomplete beta function identity (\#8.17.21 in http://dlmf.nist.gov/8.17)

\begin{equation}\label{incomplete-beta-identity}
B_x(a,b) = \frac{B(a,b)}{B(a,b+1)} \, B_x(a,b+1) - \frac{1}{b} \, x^a \, (1-x)^b
\end{equation}
with $a=\delta/2$ and $b=1/2$ gives

\begin{equation}\label{incomplete-beta-identity-particular}
B_x \left ( \frac{\delta}{2}, \frac{1}{2} \right ) x^{-\delta/2} = (\delta + 1) \, B_x \left ( \frac{\delta}{2}, \frac{3}{2} \right ) x^{-\delta/2}  - 2 \, \sqrt{1-x} \,\,\, .
\end{equation}
From this we see that the expressions~(\ref{q-nd-de-dz}) and~(\ref{q-nd}) are equivalent.  Furthermore, the latter method can be used even where there is no one-to-one correspondence between $E$ and $z$, as in the diffusional transport case, next to be considered.

\subsection{Diffusional Model}

Using Equation~(\ref{fez-result}) for the differential particle flux spectrum $F(E,z)$, the energy flux ${\cal F}(z)$ in the diffusional transport model becomes

\begin{eqnarray}\label{f-diff}
{\cal F}(z) & = & \int_0^\infty E \, F(E,z) \, dE = \sqrt{\frac{3}{\pi}} \, \frac{\dot{N}}{A} \,
\frac{(\delta-1)}{E_{c}} \times \nonumber \\
& \times & \left [ \! \left [ \int_{E = 0}^{E_c} E^2 dE \int_{E_0 = E_c}^{\infty} dE_{0} \, \frac{(E_{0}/E_{c})^{-\delta}}{(E_0^4-E^4)^{1/2}} \, \exp \left \{ - \frac{3 (Knz)^2}{E_0^4-E^4} \right \} + \right . \right . \nonumber \\
& \qquad & + \left . \left . \int_{E = E_c}^\infty E^2 dE \int_{E_0 = E}^{\infty} dE_{0} \, \frac{(E_{0}/E_{c})^{-\delta}}{(E_0^4-E^4)^{1/2}} \, \exp \left \{ - \frac{3 (Knz)^2}{E_0^4-E^4} \right \} \right ] \! \right ] \,\,\, .
\end{eqnarray}
Reversing the order of $(E,E_0)$ integration gives

\begin{equation}\label{f-diff-reverse}
{\cal F}(z) = \sqrt{\frac{3}{\pi}} \, \frac{\dot{N}}{A} \,
\frac{(\delta-1)}{E_{c}} \times \int_{E_0 = E_c}^\infty \left ( \frac{E_0}{E_c} \right )^{-\delta} \, dE_0 \, \int_{E = 0}^{E_0} \frac{E^2 \, dE}{(E_0^4-E^4)^{1/2}} \, \exp \left \{ - \frac{3 (Knz)^2}{E_0^4-E^4} \right \}  \,\,\, .
\end{equation}
From this, it is now straightforward to calculate the heating rate

\begin{eqnarray}\label{q-diff}
Q(z) & = & - \frac{d{\cal F}}{dz} = \sqrt{\frac{3}{\pi}} \, \frac{\dot{N}}{A} \,
\frac{(\delta-1)}{E_{c}} \, \left ( 6 K^2n^2 z \right ) \times \nonumber \\
& \times & \int_{E_0 = E_c}^\infty \left ( \frac{E_0}{E_c} \right )^{-\delta} dE_0 \int_{E = 0}^{E_0} \frac{E^2 \, dE}{(E_0^4-E^4)^{3/2}} \, \exp \left \{ - \frac{3 (Knz)^2}{E_0^4-E^4} \right \} \,\,\, .
\end{eqnarray}

The energy flux ${\cal F}(z)$ has a maximum at $z=0$ and hence its divergence, the heating rate $Q(0) = d{\cal F}/dz \, (0) = 0$.  As the energy flux decreases with distance, a positive heating rate develops, which subsequently decreases as the energy flux (and hence its divergence) gets smaller.  The maximum value of $Q(z)$ occurs where $dQ(z)/dz=0$, i.e., where $z$ satisfies the transcendental equation

\begin{equation}\label{z-max}
6 K^2 n^2 z^2 = \frac{\int_{E_0 = E_c}^\infty \left ( \frac{E_0}{E_c} \right )^{-\delta} dE_0 \int_{E = 0}^{E_0} \frac{E^2 \, dE}{(E_0^4-E^4)^{3/2}} \, \exp \left \{ - \frac{3 (Knz)^2}{E_0^4-E^4} \right \}}{\int_{E_0 = E_c}^\infty \left ( \frac{E_0}{E_c} \right )^{-\delta} dE_0 \int_{E = 0}^{E_0} \frac{E^2 \, dE}{(E_0^4-E^4)^{5/2}} \, \exp \left \{ - \frac{3 (Knz)^2}{E_0^4-E^4} \right \}} \,\,\, ,
\end{equation}

\begin{figure}[pht]
\begin{center}
\includegraphics[width=0.45\textwidth]{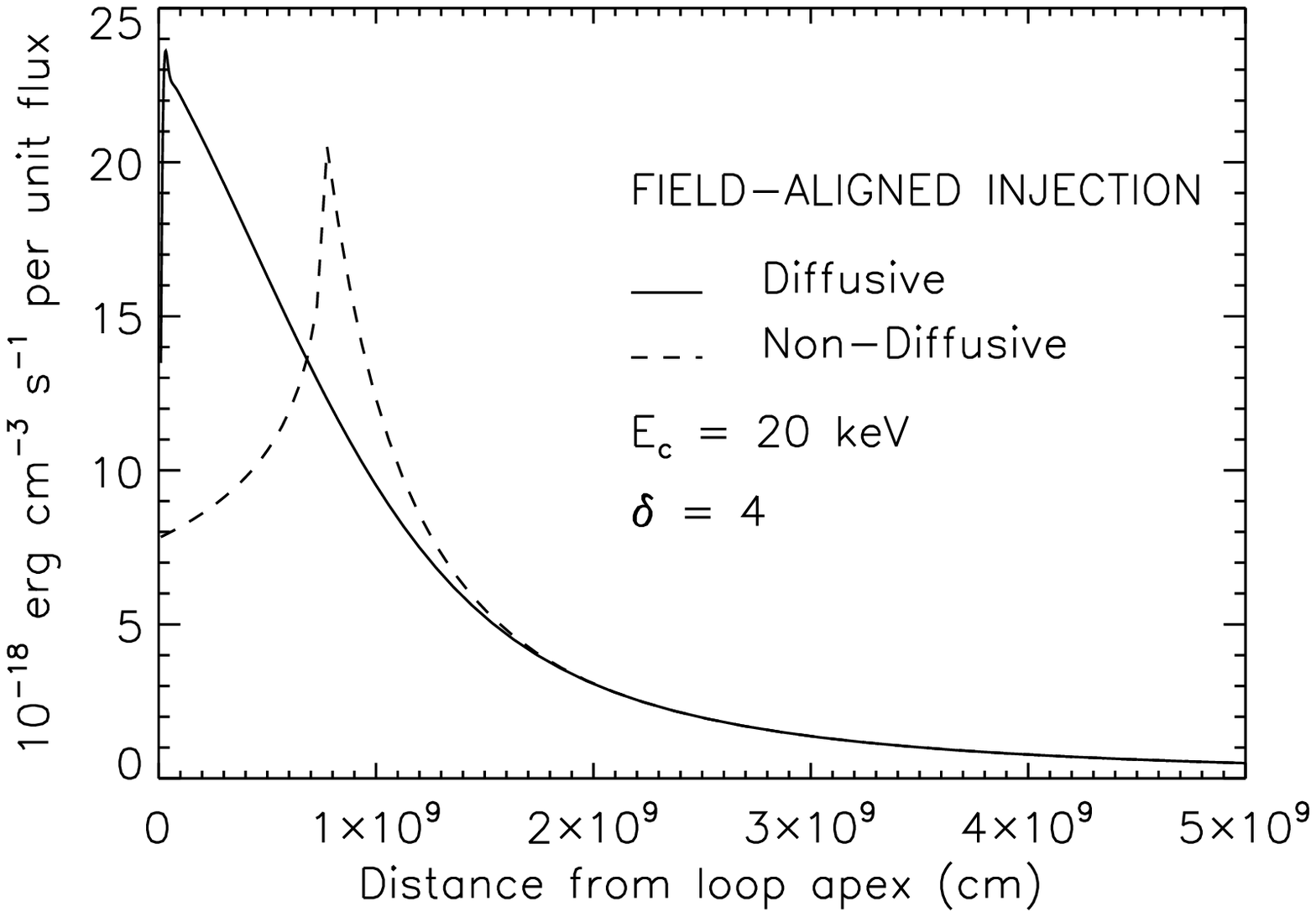}
\hspace{1cm}
\includegraphics[width=0.45\textwidth]{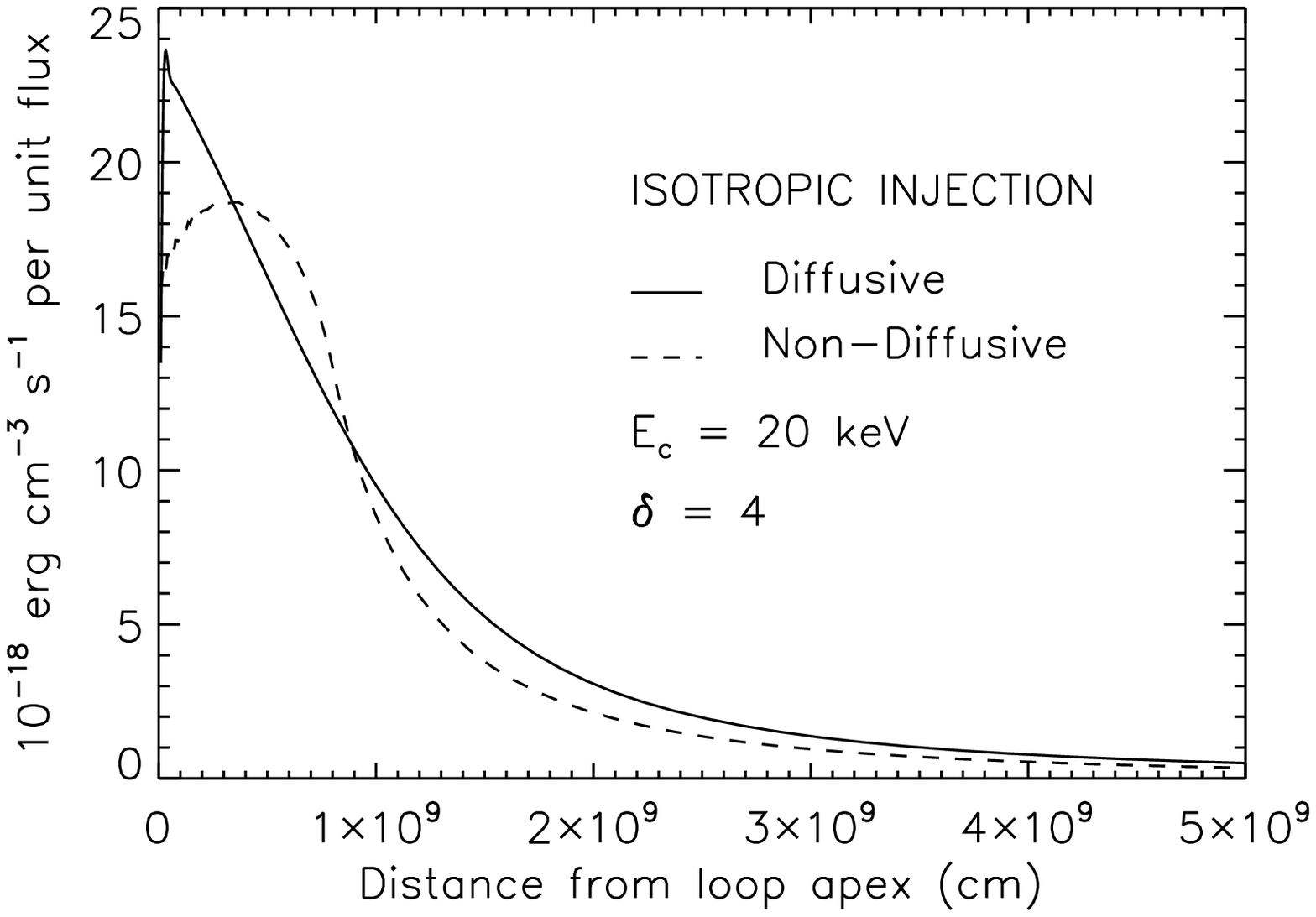}\\
\includegraphics[width=0.45\textwidth]{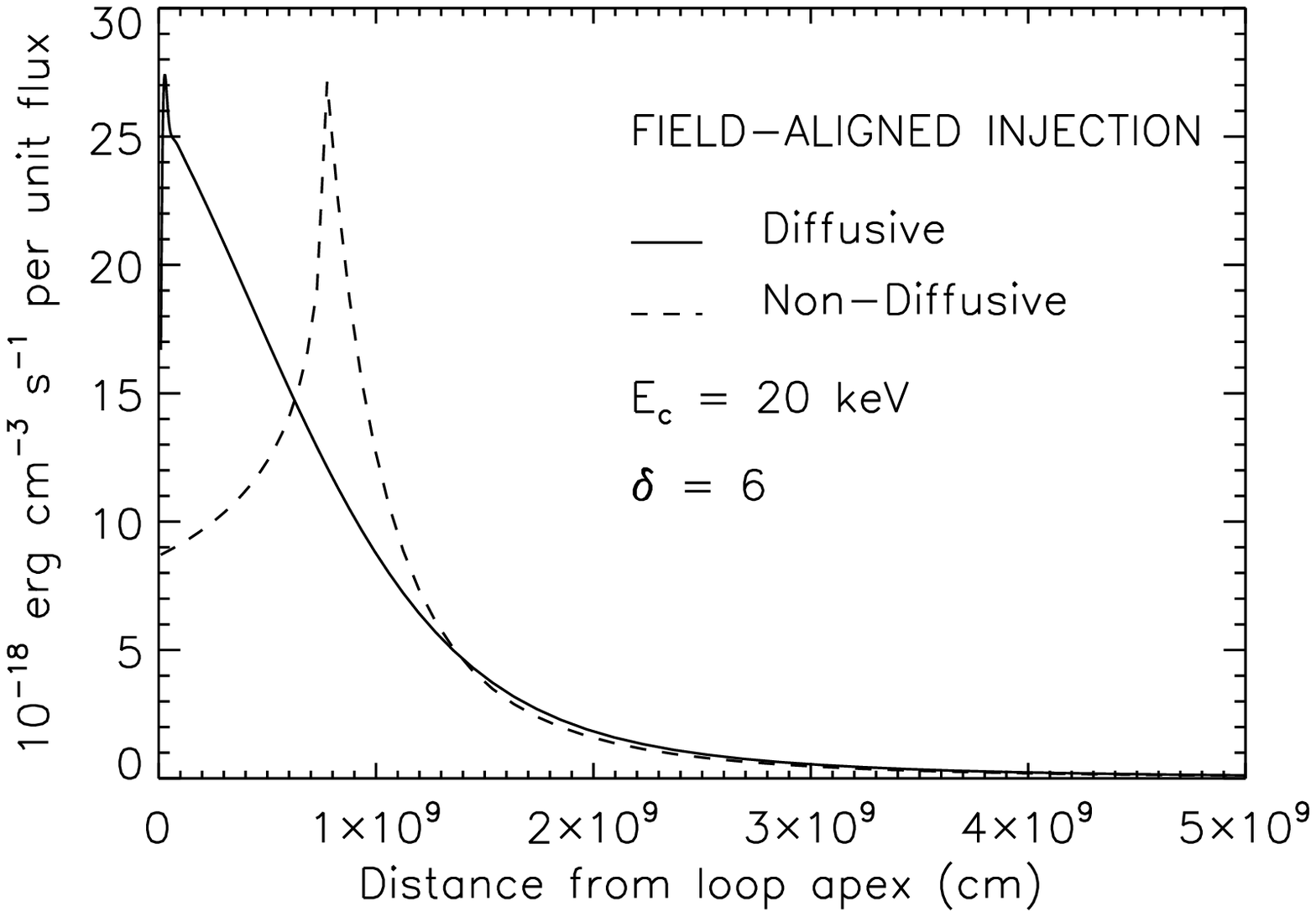}
\hspace{1cm}
\includegraphics[width=0.45\textwidth]{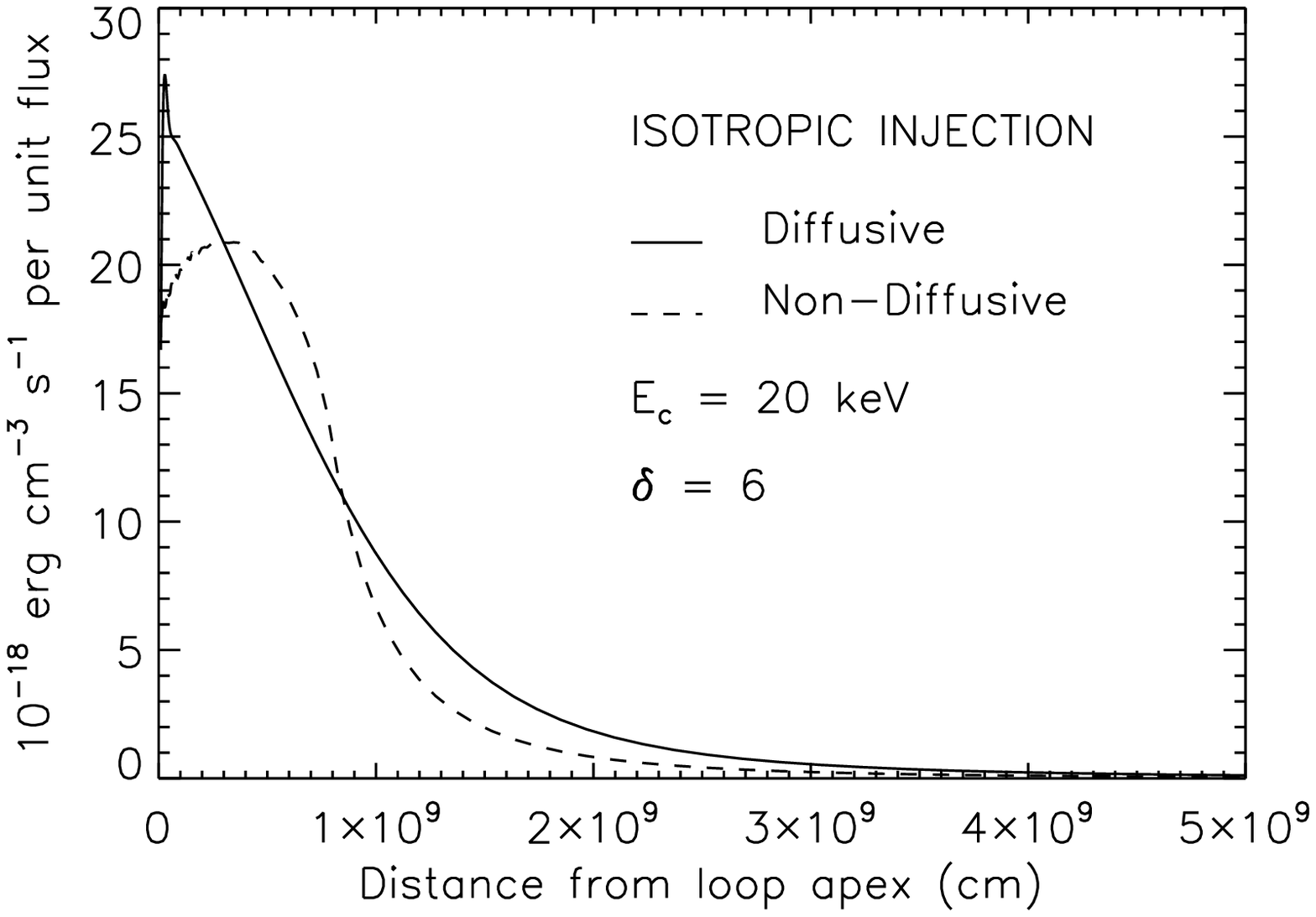}
\caption {Comparison of heating profiles for $E_c = 20$~keV and $n=10^{11}$~cm$^{-3}$. The solid line in each panel represents the heating in the diffusional propagation.  The dashed lines represent the heating in the deterministic model for a one-dimensional (field-aligned injection) model (left panel) and for a model with isotropic injection in a hemisphere (right panel).}\label{fig:comparison-ec20}
\end{center}
\end{figure}

The left-hand panels of Figure~\ref{fig:comparison-ec20} compare the heating rate~(\ref{q-nd}) in the one-dimensional deterministic model with that in the diffusional propagation model (Equation~(\ref{q-diff})).  Results are shown for $n = 10^{11}$~cm$^{-3}$ and $E_c = 20$~keV (results for different values of $n$ and $E_c$ scale and shift straightforwardly), and for $\delta=4$ and $\delta=6$ (top and bottom panels, respectively).  The right-hand panels of Figure~\ref{fig:comparison-ec20} compare the heating rate~(\ref{qnd-general-isotropic}) in a deterministic model with isotropic injection (over the downward hemisphere) with that for the diffusional propagation model (Equation~(\ref{q-diff})).  While the heating rates in all three models are of comparable magnitude, the following should be noted:

\begin{itemize}

\item the deterministic model with field-aligned injection significantly underestimates the heating near the injection point because it neglects electrons that scatter to high pitch angles and hence remain close to the injection site.  It also overestimates the heating at moderate distances, with a spike\footnote{The sharpness of this spike is somewhat artificial as it is produced by the assumed abrupt cutoff in the injected electron distribution below $E_c$.  A more gradual tapering of the injected spectrum at low energies will smooth this out; however, there will still be a (broader) peak in the heating around the locations where electrons at the spectral break point thermalize.} at distances close to where electrons of energy $E_c$ thermalize.

\item the maximum heating rate occurs at different positions in the deterministic and diffusional models, but is of comparable magnitude.

\item the results for the deterministic model with isotropic-injection in the downward hemisphere are only slightly different from the diffusional model (that involves isotropic injection over the entire sphere).  {\it This close agreement implies that the chromospheric heating rate can in most cases be adequately modeled by a deterministic transport model with isotropic injection in the downward hemisphere}.

\end{itemize}

\section{Return current Ohmic energy deposition}\label{return-current-ohmic}

For an anisotropic injection of electrons (or even an isotropic injection so that electrons proceed away from the injection point in separate hemispheres), a return current is rapidly established by the thermal electrons in the target plasma in order to effect charge and current neutralization \citep[see][]{1977ApJ...218..306K,1980ApJ...235.1055E,1984ApJ...280..448S,1985ApJ...293..584H,1989SoPh..120..343L,1990A&A...234..496V,1995A&A...304..284Z,2005A&A...432.1033Z,2006ApJ...651..553Z,2008A&A...487..337B,2013ApJ...773..121C}.
Driving this return current through the finite resistivity of the ambient medium results in an Ohmic energy deposition rate

\begin{equation}
Q_{rc}(z) = j_{\parallel}(z) \cdot {\cal E}_\parallel(z) \,\,\, ,
\end{equation}
where the return current density $j_\parallel$ is

\begin{equation}
j(z) = n e \langle v_{\parallel}\rangle = e  \int_0^\infty F(E,z) \, dE \,\,\, .
\end{equation}
For a local Ohm's law ${\cal E}_\parallel = \eta j_\parallel$, with scalar resistivity $\eta$, we thus have

\begin{equation}\label{q-rc-exp}
Q_{rc}(z) = \eta \, e^{2} \left ( \int_0^\infty F(E,z) \, dE  \right )^{2} \,\,\, .
\end{equation}
The form of $F(E,z)$ in this expression should, of course, be evaluated (or computed) self-consistently using both collisional and return-current losses.  However, as a first approximation, we can use the collisional diffusion result~(\ref{fez-result}) for $F(E,z)$ (this will be justified {\it a posteriori} below).  Reversing the order of $(E,E_0)$ integration, we obtain

\begin{eqnarray}
\int_0^\infty F(E,z) \, dE & = & \sqrt{\frac{3}{\pi}} \, \frac{\dot{N}}{A} \, \frac{(\delta-1)}{E_c} \, \times \cr & \times & \int_{E_0=E_c}^\infty \left ( \frac{E_0}{E_c} \right )^{-\delta} \, dE_0 \, \int_{E=0}^{E_0} \frac{E \, dE}{(E_0^4-E^4)^{1/2}} \exp \left \{ - \frac{3 (Knz)^2}{E_0^4-E^4} \right \} \,\,\, ,
\end{eqnarray}
so that

\begin{eqnarray}\label{return-current-diffusion-result}
Q_{rc}(z) & = & \eta \, e^2 \, \left ( \frac{3}{\pi} \right ) \, (\delta-1)^2 \, \left ( \frac{{\dot N}}{A} \right )^2 \, \times \cr
& \times & \left [ \frac{1}{E_c} \, \int_{E_0=E_c}^\infty \left ( \frac{E_0}{E_c} \right )^{-\delta} \, dE_0 \, \int_0^{E_0} \frac{E \, dE}{(E_0^4-E^4)^{1/2}} \exp \left \{ - \frac{3 (Knz)^2}{E_0^4-E^4} \right \} \right ]^2 \,\,\, .
\end{eqnarray}

The corresponding deterministic (non-diffusive) field-aligned injection result \citep[e.g.][]{1977ApJ...218..306K,1980ApJ...235.1055E} is obtained by using the form~(\ref{fez-old-model}) for $F(E,z)$ in Equation~(\ref{q-rc-exp}):

\begin{figure}[pht]
\centering
\includegraphics[width=0.7\textwidth]{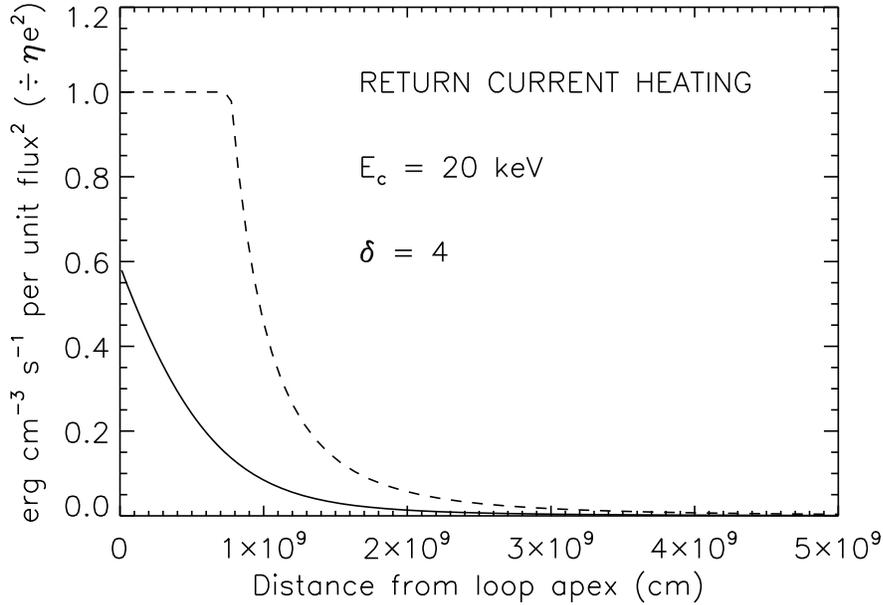}
\caption {Comparison of return current Ohmic heating profiles for $E_c = 20$~keV and $\delta =4$. The solid line represents the heating in the diffusional propagation, while the dashed line represents the heating in the deterministic field-aligned injection model. The units of the heating are per unit squared injected particle flux $({\dot N}/A)^2$, and are also scaled by the quantity $\eta \, e^2$.}\label{fig:return-current}
\end{figure}

\begin{eqnarray}
\int_0^\infty F(E,z) \, dE & = & \frac{\dot{N}}{A} \,
(\delta-1) \, E_c^{\delta-1} \, \times
\begin{cases}
\int_{\sqrt{E_c^2-2Knz}}^\infty \frac{E \, dE}{\left ( E^2 + 2Knz \right )^{(\delta+1)/2}} \, ; \quad z < \frac{E_c^2}{2Kn} \\
\int_0^\infty \frac{E \, dE}{( E^2 + 2Knz)^{(\delta+1)/2}} \qquad \quad \,\, ; \quad z > \frac{E_c^2}{2Kn}
\end{cases} \cr
& = & \frac{\dot{N}}{A} \times
\begin{cases}
1 \qquad \qquad \qquad ; \quad z < \frac{E_c^2}{2Kn} \\
\left ( \frac{2Knz}{E_c^2} \right )^{\frac{1-\delta}{2}} \quad \, ; \quad z > \frac{E_c^2}{2Kn} \,\,\, ,
\end{cases}
\end{eqnarray}
which simply reflects the conservation of particle flux down to depth $z=E_c^2/2Kn$, after which electrons are progressively ``lost'' from the beam.  Thus, for such a non-diffusive field-aligned injection model,

\begin{equation}\label{qrc-nd}
Q_{rc,ND}(z) = \eta \, e^{2} \, \left ( \frac{{\dot N}}{A} \right )^2 \, \times
\begin{cases}
1 \qquad \qquad \quad ; \quad z < \frac{E_c^2}{2Kn} \\
\left ( \frac{2Knz}{E_c^2} \right )^{1-\delta} \quad ; \quad z > \frac{E_c^2}{2Kn} \,\,\, .
\end{cases}
\end{equation}

Overall, the effect of diffusion is to reduce the anisotropy in the electron phase-space distribution function and thus reduce the magnitude of the return current and in turn the amount of Ohmic heating. Figure~\ref{fig:return-current} compares the Ohmic heating profiles (in units of $\eta e^2 (\dot{N}/A)^2$) in the diffusive and field-aligned deterministic models. Including diffusion reduces the return current heating rate by a factor of about two to three near the injection point, and by over an order of magnitude near the point where electrons at the cutoff energy $E_c$ start to be lost from the beam.

We can now justify {\it a posteriori} the use of the collision-dominated expression for $F(E,z)$ in the calculation of the return current heating rate.  An upper limit to the maximum return current heating rate is obtained by setting $z=0$ in Equation~(\ref{qrc-nd}):

\begin{equation}\label{q-max-rc}
Q_{rc,max} = \eta \, e^2 \,  \left ( \frac{{\dot N}}{A} \right )^2 \,\,\, .
\end{equation}
To compare this with the maximum heating rate in the collisional model, we use the result~(\ref{q-nd-de-dz}) for the deterministic model at $z=E_c^2/2Kn$, since Figure~\ref{fig:comparison-ec20} shows that the maximum heating rate in the diffusional model is similar.  This allows us to calculate the ratio of the maximum return current Ohmic heating to collisional heating:

\begin{equation}\label{max-heating-ratio}
\frac{Q_{rc,max}}{Q_{c,max}} = \frac{2}{(\delta - 1) \, B(\frac{\delta}{2}, \frac{1}{2})} \, \eta \, e^2 \, \frac{E_c}{Kn} \left ( \frac{{\dot N}}{A} \right ) \,\,\, .
\end{equation}
Although electron transport properties such as thermal conductivity and resistivity can be altered in the presence of additional non-collisional processes, e.g., angular scattering off, for example, magnetic inhomogeneities \citep[e.g.,][]{2016ApJ...824...78B}, for consistency with the assumed collision-dominated transport we use the \citep{1962pfig.book.....S} expression

\begin{equation}\label{eta-def}
\eta = \frac{\pi e^2 m^{1/2} \ln \Lambda}{(k_B T)^{3/2}}
\end{equation}
for the resistivity $\eta$.  With this, Equation~(\ref{max-heating-ratio}) becomes

\begin{equation}\label{max-heating-ratio-2}
\frac{Q_{rc,max}}{Q_{c,max}} = \frac{1}{(\delta-1) \, B(\frac{\delta}{2}, \frac{1}{2})} \frac{m^{1/2}}{(k_B T)^{3/2}} \, \frac{E_c}{n} \left ( \frac{{\dot N}}{A} \right ) = \frac{m^{1/2}}{8 \, (k_B T)^{3/2}} \, \frac{E_c}{n} \left ( \frac{{\dot N}}{A} \right )\,\,\, ,
\end{equation}
where we have set $\delta =4$.  Substituting $E_c = 20$~keV $=3.2 \times 10^{-8}$~erg, $T=10^7$~K, and $n=10^{11}$~cm$^{-3}$ gives

\begin{equation}\label{max-heating-ratio-3}
\frac{Q_{rc,max}}{Q_{c,max}} \simeq 2.4 \times 10^{-20} \, \left ( \frac{{\dot N}}{A} \right ) \,\,\, .
\end{equation}
Even for a large flare with ${\dot N} = 10^{37}$~s$^{-1}$ and $A=10^{18}$~cm$^2$, this gives $Q_{rc,max}/Q_{c,max} \simeq 1/4$.  This ratio is even smaller in the diffusional model: although the maximum collisional heating rates in the diffusional and deterministic models are comparable (Figure~\ref{fig:comparison-ec20}), return current losses are significantly reduced relative to those in the deterministic model (Figure~\ref{fig:return-current}). We therefore see that return current ohmic losses are significantly less than collisional losses, so that the evolution of $F(E,z)$ is controlled primarily by collisions.  Thus the use of a collisional form for $F(E,z)$ in determining the approximate return current losses is justified {\it a posteriori}.

\section{Summary and Conclusions}\label{summary-conclusions}

Modelling of energy deposition by injected electron beams in solar flares previously assumed a directional beam accelerated in a point source and directed downward to the chromosphere where it was stopped collisionally. However, the need to include angular diffusion due to collisions in the physics of electron transport \citep{2017ApJ...835..262B} results in significantly changed profiles for the electron flux versus depth and hence for the profile $Q(z)$ of heat deposition versus depth. The resulting cold target heating function can, however be adequately modeled simply by using a deterministic transport model with isotropic injection in the downward hemisphere (see right panels of Figure~\ref{fig:comparison-ec20}).  The effects on Ohmic return current heating are more severe; the significantly greater level of isotropization of the injected electrons caused by enhanced pitch angle scattering reduces the magnitude of the associated current, resulting in a reduction of up to an order-of-magnitude in the Ohmic heating rate associated with the neutralizing return current.

This treatment can be extended to include non-collisional pitch-angle scattering
of electrons in flaring loops \citep{2014ApJ...780..176K,2018A&A...610A...6M}.
In a future work we will use these modified heating functions to determine the hydrodynamic response \citep{2015ApJ...809..104A} of the solar atmosphere to the electron energy input.  This will in turn allow us to construct velocity differential emission measure \citep{1995ApJ...447..915N} profiles with which to compare observations of shifted and broadened soft X-ray and EUV spectral lines, with the ultimate goal of more meaningfully constraining the processes of nonthermal electron acceleration and transport during solar flares.

\acknowledgments NHB and AGE were supported by grant NNX17AI16G from NASA's Heliophysics Supporting Research program. EPK was supported by a STFC consolidated grant ST/P000533/1.

\bibliographystyle{apj}
\bibliography{diffusive_energy_deposition_refs}

\begin{thebibliography}{56}
\expandafter\ifx\csname natexlab\endcsname\relax\def\natexlab#1{#1}\fi

\bibitem[{{Allred} {et~al.}(2015){Allred}, {Kowalski}, \&
  {Carlsson}}]{2015ApJ...809..104A}
{Allred}, J.~C., {Kowalski}, A.~F., \& {Carlsson}, M. 2015, \apj, 809, 104

\bibitem[{{Antonucci} {et~al.}(1982){Antonucci}, {Gabriel}, {Acton},
  {Leibacher}, {Culhane}, {Rapley}, {Doyle}, {Machado}, \&
  {Orwig}}]{1982SoPh...78..107A}
{Antonucci}, E., {Gabriel}, A.~H., {Acton}, L.~W., {Leibacher}, J.~W.,
  {Culhane}, J.~L., {Rapley}, C.~G., {Doyle}, J.~G., {Machado}, M.~E., \&
  {Orwig}, L.~E. 1982, \solphys, 78, 107

\bibitem[{{Battaglia} \& {Benz}(2008)}]{2008A&A...487..337B}
{Battaglia}, M., \& {Benz}, A.~O. 2008, \aap, 487, 337

\bibitem[{{Bentley} {et~al.}(1994){Bentley}, {Doschek}, {Simnett}, {Rilee},
  {Mariska}, {Culhane}, {Kosugi}, \& {Watanabe}}]{1994ApJ...421L..55B}
{Bentley}, R.~D., {Doschek}, G.~A., {Simnett}, G.~M., {Rilee}, M.~L.,
  {Mariska}, J.~T., {Culhane}, J.~L., {Kosugi}, T., \& {Watanabe}, T. 1994,
  \apjl, 421, L55

\bibitem[{{Bespalov} {et~al.}(1991){Bespalov}, {Zaitsev}, \&
  {Stepanov}}]{1991ApJ...374..369B}
{Bespalov}, P.~A., {Zaitsev}, V.~V., \& {Stepanov}, A.~V. 1991, \apj, 374, 369

\bibitem[{{Bian} {et~al.}(2017){Bian}, {Emslie}, \&
  {Kontar}}]{2017ApJ...835..262B}
{Bian}, N.~H., {Emslie}, A.~G., \& {Kontar}, E.~P. 2017, \apj, 835, 262

\bibitem[{{Bian} {et~al.}(2016){Bian}, {Kontar}, \&
  {Emslie}}]{2016ApJ...824...78B}
{Bian}, N.~H., {Kontar}, E.~P., \& {Emslie}, A.~G. 2016, \apj, 824, 78

\bibitem[{{Bian} {et~al.}(2011){Bian}, {Kontar}, \&
  {MacKinnon}}]{2011A&A...535A..18B}
{Bian}, N.~H., {Kontar}, E.~P., \& {MacKinnon}, A.~L. 2011, \aap, 535, A18

\bibitem[{{Brosius} {et~al.}(2016){Brosius}, {Daw}, \&
  {Inglis}}]{2016ApJ...830..101B}
{Brosius}, J.~W., {Daw}, A.~N., \& {Inglis}, A.~R. 2016, \apj, 830, 101

\bibitem[{{Brosius} \& {Inglis}(2017)}]{2017ApJ...848...39B}
{Brosius}, J.~W., \& {Inglis}, A.~R. 2017, \apj, 848, 39

\bibitem[{{Brosius} \& {Phillips}(2004)}]{2004ApJ...613..580B}
{Brosius}, J.~W., \& {Phillips}, K.~J.~H. 2004, \apj, 613, 580

\bibitem[{{Brown}(1971)}]{1971SoPh...18..489B}
{Brown}, J.~C. 1971, \solphys, 18, 489

\bibitem[{{Brown}(1972)}]{1972SoPh...26..441B}
---. 1972, \solphys, 26, 441

\bibitem[{{Brown}(1973)}]{1973SoPh...31..143B}
---. 1973, \solphys, 31, 143

\bibitem[{{Codispoti} {et~al.}(2013){Codispoti}, {Torre}, {Piana}, \&
  {Pinamonti}}]{2013ApJ...773..121C}
{Codispoti}, A., {Torre}, G., {Piana}, M., \& {Pinamonti}, N. 2013, \apj, 773,
  121

\bibitem[{{Doschek}(1990)}]{1990ApJS...73..117D}
{Doschek}, G.~A. 1990, \apjs, 73, 117

\bibitem[{{Emslie}(1978)}]{1978ApJ...224..241E}
{Emslie}, A.~G. 1978, \apj, 224, 241

\bibitem[{{Emslie}(1980)}]{1980ApJ...235.1055E}
---. 1980, \apj, 235, 1055

\bibitem[{{Emslie} {et~al.}(2012){Emslie}, {Dennis}, {Shih}, {Chamberlin},
  {Mewaldt}, {Moore}, {Share}, {Vourlidas}, \& {Welsch}}]{2012ApJ...759...71E}
{Emslie}, A.~G., {Dennis}, B.~R., {Shih}, A.~Y., {Chamberlin}, P.~C.,
  {Mewaldt}, R.~A., {Moore}, C.~S., {Share}, G.~H., {Vourlidas}, A., \&
  {Welsch}, B.~T. 2012, \apj, 759, 71

\bibitem[{{Emslie} \& {Smith}(1984)}]{1984ApJ...279..882E}
{Emslie}, A.~G., \& {Smith}, D.~F. 1984, \apj, 279, 882

\bibitem[{{Emslie} \& {Alexander}(1987)}]{1987SoPh..110..295E}
{Emslie}, G.~A., \& {Alexander}, D. 1987, \solphys, 110, 295

\bibitem[{{G{\"o}m{\"o}ry} {et~al.}(2016){G{\"o}m{\"o}ry}, {Veronig}, {Su},
  {Temmer}, \& {Thalmann}}]{2016A&A...588A...6G}
{G{\"o}m{\"o}ry}, P., {Veronig}, A.~M., {Su}, Y., {Temmer}, M., \& {Thalmann},
  J.~K. 2016, \aap, 588, A6

\bibitem[{{Holman}(1985)}]{1985ApJ...293..584H}
{Holman}, G.~D. 1985, \apj, 293, 584

\bibitem[{{Holman} {et~al.}(2011){Holman}, {Aschwanden}, {Aurass}, {Battaglia},
  {Grigis}, {Kontar}, {Liu}, {Saint-Hilaire}, \&
  {Zharkova}}]{2011SSRv..159..107H}
{Holman}, G.~D., {Aschwanden}, M.~J., {Aurass}, H., {Battaglia}, M., {Grigis},
  P.~C., {Kontar}, E.~P., {Liu}, W., {Saint-Hilaire}, P., \& {Zharkova}, V.~V.
  2011, \ssr, 159, 107

\bibitem[{{Jeffrey} {et~al.}(2014){Jeffrey}, {Kontar}, {Bian}, \&
  {Emslie}}]{2014ApJ...787...86J}
{Jeffrey}, N.~L.~S., {Kontar}, E.~P., {Bian}, N.~H., \& {Emslie}, A.~G. 2014,
  \apj, 787, 86

\bibitem[{{Knight} \& {Sturrock}(1977)}]{1977ApJ...218..306K}
{Knight}, J.~W., \& {Sturrock}, P.~A. 1977, \apj, 218, 306

\bibitem[{{Kontar} {et~al.}(2014){Kontar}, {Bian}, {Emslie}, \&
  {Vilmer}}]{2014ApJ...780..176K}
{Kontar}, E.~P., {Bian}, N.~H., {Emslie}, A.~G., \& {Vilmer}, N. 2014, \apj,
  780, 176

\bibitem[{{Kontar} {et~al.}(2011{\natexlab{a}}){Kontar}, {Brown}, {Emslie},
  {Hajdas}, {Holman}, {Hurford}, {Ka{\v s}parov{\'a}}, {Mallik}, {Massone},
  {McConnell}, {Piana}, {Prato}, {Schmahl}, \&
  {Suarez-Garcia}}]{2011SSRv..159..301K}
{Kontar}, E.~P., {Brown}, J.~C., {Emslie}, A.~G., {Hajdas}, W., {Holman},
  G.~D., {Hurford}, G.~J., {Ka{\v s}parov{\'a}}, J., {Mallik}, P.~C.~V.,
  {Massone}, A.~M., {McConnell}, M.~L., {Piana}, M., {Prato}, M., {Schmahl},
  E.~J., \& {Suarez-Garcia}, E. 2011{\natexlab{a}}, \ssr, 159, 301

\bibitem[{{Kontar} {et~al.}(2011{\natexlab{b}}){Kontar}, {Hannah}, \&
  {Bian}}]{2011ApJ...730L..22K}
{Kontar}, E.~P., {Hannah}, I.~G., \& {Bian}, N.~H. 2011{\natexlab{b}}, \apjl,
  730, L22

\bibitem[{{Kontar} {et~al.}(2017){Kontar}, {Perez}, {Harra}, {Kuznetsov},
  {Emslie}, {Jeffrey}, {Bian}, \& {Dennis}}]{2017PhRvL.118o5101K}
{Kontar}, E.~P., {Perez}, J.~E., {Harra}, L.~K., {Kuznetsov}, A.~A., {Emslie},
  A.~G., {Jeffrey}, N.~L.~S., {Bian}, N.~H., \& {Dennis}, B.~R. 2017, Physical
  Review Letters, 118, 155101

\bibitem[{{Landi} {et~al.}(2003){Landi}, {Feldman}, {Innes}, \&
  {Curdt}}]{2003ApJ...582..506L}
{Landi}, E., {Feldman}, U., {Innes}, D.~E., \& {Curdt}, W. 2003, \apj, 582, 506

\bibitem[{{Larosa} \& {Emslie}(1989)}]{1989SoPh..120..343L}
{Larosa}, T.~N., \& {Emslie}, A.~G. 1989, \solphys, 120, 343

\bibitem[{{Leach} \& {Petrosian}(1981)}]{1981ApJ...251..781L}
{Leach}, J., \& {Petrosian}, V. 1981, \apj, 251, 781

\bibitem[{{Li} {et~al.}(2017){Li}, {Ning}, {Huang}, \&
  {Zhang}}]{2017ApJ...841L...9L}
{Li}, D., {Ning}, Z.~J., {Huang}, Y., \& {Zhang}, Q.~M. 2017, \apjl, 841, L9

\bibitem[{{Li} {et~al.}(2015{\natexlab{a}}){Li}, {Ning}, \&
  {Zhang}}]{2015ApJ...813...59L}
{Li}, D., {Ning}, Z.~J., \& {Zhang}, Q.~M. 2015{\natexlab{a}}, \apj, 813, 59

\bibitem[{{Li} {et~al.}(2015{\natexlab{b}}){Li}, {Ding}, {Qiu}, \&
  {Cheng}}]{2015ApJ...811....7L}
{Li}, Y., {Ding}, M.~D., {Qiu}, J., \& {Cheng}, J.~X. 2015{\natexlab{b}}, \apj,
  811, 7

\bibitem[{{Lin} \& {Hudson}(1976)}]{1976SoPh...50..153L}
{Lin}, R.~P., \& {Hudson}, H.~S. 1976, \solphys, 50, 153

\bibitem[{{MacKinnon} \& {Craig}(1991)}]{1991A&A...251..693M}
{MacKinnon}, A.~L., \& {Craig}, I.~J.~D. 1991, \aap, 251, 693

\bibitem[{{Mariska} {et~al.}(1993){Mariska}, {Doschek}, \&
  {Bentley}}]{1993ApJ...419..418M}
{Mariska}, J.~T., {Doschek}, G.~A., \& {Bentley}, R.~D. 1993, \apj, 419, 418

\bibitem[{{Mariska} {et~al.}(1989){Mariska}, {Emslie}, \&
  {Li}}]{1989ApJ...341.1067M}
{Mariska}, J.~T., {Emslie}, A.~G., \& {Li}, P. 1989, \apj, 341, 1067

\bibitem[{{Milligan}(2015)}]{2015SoPh..290.3399M}
{Milligan}, R.~O. 2015, \solphys, 290, 3399

\bibitem[{{Musset} {et~al.}(2018){Musset}, {Kontar}, \&
  {Vilmer}}]{2018A&A...610A...6M}
{Musset}, S., {Kontar}, E.~P., \& {Vilmer}, N. 2018, \aap, 610, A6

\bibitem[{{Newton} {et~al.}(1995){Newton}, {Emslie}, \&
  {Mariska}}]{1995ApJ...447..915N}
{Newton}, E.~K., {Emslie}, A.~G., \& {Mariska}, J.~T. 1995, \apj, 447, 915

\bibitem[{{Polito} {et~al.}(2016){Polito}, {Reep}, {Reeves}, {Sim{\~o}es},
  {Dud{\'{\i}}k}, {Del Zanna}, {Mason}, \& {Golub}}]{2016ApJ...816...89P}
{Polito}, V., {Reep}, J.~W., {Reeves}, K.~K., {Sim{\~o}es}, P.~J.~A.,
  {Dud{\'{\i}}k}, J., {Del Zanna}, G., {Mason}, H.~E., \& {Golub}, L. 2016,
  \apj, 816, 89

\bibitem[{{Reep} {et~al.}(2015){Reep}, {Bradshaw}, \&
  {Alexander}}]{2015ApJ...808..177R}
{Reep}, J.~W., {Bradshaw}, S.~J., \& {Alexander}, D. 2015, \apj, 808, 177

\bibitem[{{Rilee} \& {Doschek}(2001)}]{2001ApJ...554..464R}
{Rilee}, M.~L., \& {Doschek}, G.~A. 2001, \apj, 554, 464

\bibitem[{{Spicer} \& {Sudan}(1984)}]{1984ApJ...280..448S}
{Spicer}, D.~S., \& {Sudan}, R.~N. 1984, \apj, 280, 448

\bibitem[{{Spitzer}(1962)}]{1962pfig.book.....S}
{Spitzer}, L. 1962, Physics of Fully Ionized Gases (New York: Interscience)

\bibitem[{{Tandberg-Hanssen} \& {Emslie}(1988)}]{1988psf..book.....T}
{Tandberg-Hanssen}, E., \& {Emslie}, A.~G. 1988, The physics of solar flares
  (Cambridge and New York, Cambridge University Press)

\bibitem[{{Tian} {et~al.}(2015){Tian}, {Young}, {Reeves}, {Chen}, {Liu}, \&
  {McKillop}}]{2015ApJ...811..139T}
{Tian}, H., {Young}, P.~R., {Reeves}, K.~K., {Chen}, B., {Liu}, W., \&
  {McKillop}, S. 2015, \apj, 811, 139

\bibitem[{{van den Oord}(1990)}]{1990A&A...234..496V}
{van den Oord}, G.~H.~J. 1990, \aap, 234, 496

\bibitem[{{Warren} {et~al.}(2016){Warren}, {Reep}, {Crump}, \&
  {Sim{\~o}es}}]{2016ApJ...829...35W}
{Warren}, H.~P., {Reep}, J.~W., {Crump}, N.~A., \& {Sim{\~o}es}, P.~J.~A. 2016,
  \apj, 829, 35

\bibitem[{{Zharkova} {et~al.}(2011){Zharkova}, {Arzner}, {Benz}, {Browning},
  {Dauphin}, {Emslie}, {Fletcher}, {Kontar}, {Mann}, {Onofri}, {Petrosian},
  {Turkmani}, {Vilmer}, \& {Vlahos}}]{2011SSRv..159..357Z}
{Zharkova}, V.~V., {Arzner}, K., {Benz}, A.~O., {Browning}, P., {Dauphin}, C.,
  {Emslie}, A.~G., {Fletcher}, L., {Kontar}, E.~P., {Mann}, G., {Onofri}, M.,
  {Petrosian}, V., {Turkmani}, R., {Vilmer}, N., \& {Vlahos}, L. 2011, \ssr,
  159, 357

\bibitem[{{Zharkova} {et~al.}(1995){Zharkova}, {Brown}, \&
  {Syniavskii}}]{1995A&A...304..284Z}
{Zharkova}, V.~V., {Brown}, J.~C., \& {Syniavskii}, D.~V. 1995, \aap, 304, 284

\bibitem[{{Zharkova} \& {Gordovskyy}(2005)}]{2005A&A...432.1033Z}
{Zharkova}, V.~V., \& {Gordovskyy}, M. 2005, \aap, 432, 1033

\bibitem[{{Zharkova} \& {Gordovskyy}(2006)}]{2006ApJ...651..553Z}
---. 2006, \apj, 651, 553

\end{thebibliography}

\end{document}